\documentclass[preprint]{JHEP3}

\usepackage{amsmath,amssymb,amscd,amsfonts,graphicx,bbm}
\usepackage{multirow}

\preprint{
KUNS-2068 \\
RIKEN-TH 97\\
RBRC-682\\

WIS/19/07-Nov-DPP
}

\title{Phase structure of twisted Eguchi-Kawai model}

\author{
         Tatsuo Azeyanagi$^a$, 
         Masanori Hanada$^{b,c}$, 
         Tomoyoshi Hirata$^{a}$ and 
         Tomomi Ishikawa$^{d}$\\
\llap{$^a$}
Department of Physics, Kyoto University,\\
Kyoto 606-8502, Japan\\
\llap{$^{b}$}
Theoretical Physics Laboratory, RIKEN Nishina Center,\\
Wako, Saitama 351-0198, Japan\\
\llap{$^{c}$}
Department of Particle Physics, Weizmann Institute of Science,\\
Rehovot 76100, Israel\\
\llap{$^d$}
RIKEN BNL Research Center, Brookhaven National Laboratory,\\
Upton, New York 11973, USA\\
E-mail: \email{
aze@gauge.scphys.kyoto-u.ac.jp,
masanori.hanada@weizmann.ac.il,\\
hirata@gauge.scphys.kyoto-u.ac.jp,
tomomi@quark.phy.bnl.gov}
}

\abstract{
Twisted Eguchi-Kawai model is a useful tool for studying
the large-$N$ gauge theory.
It can also provide a nonperturbative formulation of 
the gauge theory on noncommutative spaces. 
Recently it was found that the $\mathbb{Z}_N^4$ symmetry in this model,
which is crucial for the above applications,
can break spontaneously in the intermediate coupling region.
In this article, we study the phase structure of this model
using the Monte-Carlo simulation.
In particular, we elaborately investigate the symmetry breaking point
from the weak coupling side.
The simulation results show that we cannot take a continuum limit for this model. 
}


\begin{document}

\section{Introduction}
The large-$N$ gauge theories provide fruitful features to both
phenomenology and string theory.
They are simplified in the large-$N$ limit
while preserving essential features of QCD \cite{'t Hooft:1973jz}.
Additionally, dimensional reductions of ten-dimensional ${\cal N}=1$ super
Yang-Mills theory (matrix model) are expected to provide nonperturbative
formulations of superstring theory
\cite{Banks:1996vh,Ishibashi:1996xs,Dijkgraaf:1997vv,Maldacena:1997re},
and can also be regarded as effective actions of D-branes \cite{Witten:1995ex}.
Furthermore, their twisted reduced versions, which we study in this article,
can provide a nonperturbative formulation of the gauge theories
on noncommutative spaces (NCYM) \cite{Aoki:1999vr,Ambjorn:1999ts}.
In order to study the nonperturbative nature of these theories,
numerical simulations using lattice regularizations are quite efficient.
(Non-lattice simulations are also applicable for the reduced models.
See references \cite{Hanada:2007ti,Anagnostopoulos:2007fw}
for the recent progress.)

In the large-$N$ limit there is an equivalence between
the gauge theory and its zero-dimensional reduction,
which is known as {\it Eguchi-Kawai equivalence} \cite{Eguchi:1982nm}.
Here, we consider the $SU(N)$ gauge theory (YM) on $D$-dimensional periodic
lattice with the Wilson's plaquette action
\begin{equation}
S_W=-\beta N\sum_x\sum_{\mu\neq\nu}{\rm Tr}
~U_{\mu}(x)U_{\nu}(x+\hat{\mu})U^\dagger_{\mu}(x+\hat{\nu})U^\dagger_{\nu}(x),
\end{equation}
where $U_{\mu}(x)\ (\mu=1,..., D)\in SU(N)$ are link variables and
$\beta$ is the inverse of the bare 't~Hooft coupling.
In the large-$N$ limit the space-time degrees of freedom can be neglected,
and then this theory can be equivalent to a model defined on a single
hyper-cube,
\begin{equation}
S_{EK}=-\beta N\sum_{\mu\neq\nu}{\rm Tr}
~U_{\mu}U_{\nu}U^\dagger_{\mu}U^\dagger_{\nu},
\end{equation}
which is called the Eguchi-Kawai model (EK model).
The equality was shown by observing that the Schwinger-Dyson equations
for Wilson loops (loop equations) in both theories are the same.
In the EK model the loop equations can naively have open Wilson
line terms, which do not exist in the original gauge theory side
due to the gauge invariance.
Therefore we need to assume that the global $\mathbb{Z}_N^D$ symmetry
\begin{equation}
U_\mu\to e^{i\theta_\mu}U_\mu,
\end{equation}
which eliminates the non-zero expectation value of the open Wilson lines,
is not broken spontaneously. 
However, soon after the discovery of the equivalence,
it was found that the $\mathbb{Z}_N^D$ symmetry is actually broken for $D>2$
in the weak coupling region \cite{Bhanot:1982sh}.
Although the naive EK equivalence does not hold, 
modifications were proposed for this issue.
They are quenched Eguchi-Kawai model (QEK model)
\cite{Bhanot:1982sh,Parisi:1982gp,Gross:1982at}
and twisted Eguchi-Kawal model (TEK model) \cite{GonzalezArroyo:1982ub}.
Historically, most of the works previously done were based on the
TEK model because this model is theoretically interesting and
numerically more practical
(and this model describes the NCYM as mentioned before).

In the TEK model, twisted boundary conditions are imposed and then
the $\mathbb{Z}_N^D$ symmetry is ensured in the weak coupling limit.
It is not obvious whether the symmetry is broken or not
in the intermediate coupling region. 
There is no guarantee for not violating the symmetry.
Numerical simulations in the 1980s, however, suggested that
the $\mathbb{Z}_N^D$ symmetry is not broken throughout the whole
coupling region.
Then we have believed that the TEK model actually describes
the large-$N$ limit of the gauge theory.

Recently some indication about the $\mathbb{Z}_N^D$ symmetry breaking
was surprisingly reported in several context around the TEK model
\cite{IO03,Teper:2006sp}\cite{Guralnik:2002ru,Bietenholz:2006cz}.
The most relevant discussion for the present article was done
by Teper and Vairinhos in \cite{Teper:2006sp}\footnote{
In \cite{Guralnik:2002ru,Bietenholz:2006cz} a similar model 
with two commutative and two noncommutative dimensions were studied 
in the context of NCYM. In this case the instability of ${\mathbb Z}_N$ 
preserving vacuum was observed even in a perturbative calculation. 
This instability arises due to UV/IR mixing.}
. 
They showed that the $\mathbb{Z}_N^D$ symmetry is really broken
in the intermediate coupling region by the Monte-Carlo simulation
for the $D=4$ TEK model with the standard twist.
Our work in this article is along this line and we mainly concentrate
on investigating locations of the symmetry breaking
from the weak coupling side in $(\beta, N)$ plane.
By the Monte-Carlo simulation we clarify the linear behavior of critical
lattice coupling
\begin{equation}
\beta_c^L\sim L^2,  
\label{EQ:scaling_of_beta_c^L}
\end{equation}
where $\beta_c^L$ represents critical lattice coupling from the weak
coupling side and $L$ is the lattice size we have considered.
This result means that {\it the continuum limit of the planar gauge theory 
cannot be described by the TEK model}
from the
 argument of the scaling behavior around the weak coupling limit. 
This discussion can be also applied to the NCYM case.

This article is organized as follows.
In the next section we review the TEK model briefly and fix our setup.
In section \ref{SEC:breaking} we show the numerical results for the
$\mathbb{Z}_N^D$ symmetry breaking of the TEK model and find the
scaling behavior (\ref{EQ:scaling_of_beta_c^L}). 
In section \ref{SEC:discussions} we give the validation for the
numerical result, and also discuss whether the TEK model has
a continuum limit or not. 

\section{Twisted Eguchi-Kawai model}

\subsection {Action and Wilson loop}

In this study, we treat the $D=4$ case.
The TEK model \cite{GonzalezArroyo:1982ub} is a matrix model
defined by the partition function 
\begin{equation}
Z_{TEK}=\int\prod_{\mu=1}^4 dU_{\mu}\exp(-S_{TEK}),
\end{equation}
with the action
\begin{equation}
S_{TEK}=-\beta N\sum_{\mu\neq\nu}Z_{\mu\nu}{\rm Tr}
~U_\mu U_\nu U_\mu^\dagger U_\nu^\dagger,
\label{EQ:TEK_action}
\end{equation}
where $U_{\mu}$ and $dU_\mu\ (\mu=1,2,3,4)$ are link variables and
Haar measure.
The phase factors $Z_{\mu\nu}$ are
\begin{equation}
Z_{\mu\nu}=\exp\left(2\pi i n_{\mu\nu}/N\right),\qquad
n_{\mu\nu}=-n_{\nu\mu}\in \mathbb{Z}_N.
\end{equation}
The Wilson loop operator also contains the phase $Z({\cal C})$ as
\begin{equation}
W_{TEK}({\cal C})\equiv Z({\cal C})\langle\hat{W}({\cal C})\rangle,
\end{equation}
where $\hat{W}({\cal C})$ is the trace of the product of link variables
along a contour $\cal C$ and $Z({\cal C})$ is the product of $Z_{\mu\nu}$'s
which correspond to the plaquettes in a surface whose boundary is $\cal C$.
This model is obtained by dimensional reduction of the Wilson's lattice
gauge theory with the twisted boundary condition.
With these definitions, the loop equations in the TEK model take the same
form as those in the ordinary lattice gauge theory if the $\mathbb{Z}_N^4$
symmetry, which we discuss in section \ref{SEC:U(1)symmetry}, is not broken. 

\subsection {Twist prescriptions and classical solutions}
\label{SEC:twist prescription}

In the weak coupling limit, the path-integral is dominated by
the configuration which gives the minimum to the action.
This configuration $U^{(0)}_\mu=\Gamma_{\mu}$ satisfies the 't~Hooft algebra
\begin{equation}
\Gamma_{\mu}\Gamma_{\nu}=Z_{\nu\mu}\Gamma_{\nu}\Gamma_{\mu},
\end{equation}
and is called ``twist-eater''.
The most popular twist might be the minimal symmetric twist (standard twist)
\begin{eqnarray}
n_{\mu\nu}&=&\left(\begin{array}{cccc}
 0 &  L &  L & L\\
-L &  0 &  L & L\\
-L & -L &  0 & L\\
-L & -L & -L & 0
\end{array}\right),\qquad N=L^2.
\label{EQ:minimal_sym_form}
\end{eqnarray}
This twist represents $L^4$ lattice.
In order to construct the classical solution for this twist,
it is convenient to use the $SL(4, \mathbb{Z})$ transformation for the
coordinates on $\mathbb{T}^4$ \cite{vanBaal:1985na}.
Using the $SL(4, \mathbb{Z})$ transformation we can always rewrite
the $n_{\mu\nu}$ in the skew-diagonal form
\begin{eqnarray}
n_{\mu\nu}&\longrightarrow&n_{\mu\nu}'=V^T n_{\mu\nu}V=
\left(\begin{array}{cc|cc}
 0 &  L &  0 & 0\\
-L &  0 &  0 & 0\\
\hline
 0 &  0 &  0 & L\\
 0 &  0 & -L & 0
\end{array}\right),
\label{EQ:minimal_skew_form}
\end{eqnarray}
where $V$ is a $SL(4, \mathbb{Z})$ transformation matrix.
This form makes the construction of the twist-eater easy.
Here we define $L\times L$ ``shift'' matrix $\hat{S}_L$ and ``clock''
matrix $\hat{C}_L$ by
\begin{eqnarray}
\hat{S}_L=\left(
\begin{array}{ccccc}
0      & ~1     & 0      & ~\cdots & ~0     \\
\vdots & ~0     & 1      & ~\ddots & \vdots \\
\vdots &        & \ddots & ~\ddots & ~0     \\
0      &        &        & ~\ddots & ~1     \\
1      & ~0     & \cdots & ~\cdots & ~0
\end{array}
\right), \qquad
\hat{C}_L=\left(
\begin{array}{ccccc}
1 &              & & & \multirow{2}{0mm}[-1mm]{\it\huge O}         \\
 & e^{2\pi i/L} &                     &        &                   \\
 &              & e^{2\pi i\cdot 2/L} &        &                   \\
 &              &                     & \ddots &                   \\
\multirow{2}{0mm}[4mm]{\it\huge O} & & &        & e^{2\pi i(L-1)/L}
\end{array}
\right),
\end{eqnarray}
which satisfy the little 't~Hooft algebra
\begin{equation}
\hat{C}_L\hat{S}_L =  e^{-2\pi i/L}\hat{S}_L\hat{C}_L.
\end{equation}
Using these matrices, the twist-eater configuration for the skew-diagonal
form (\ref{EQ:minimal_skew_form}) is easily constructed as
\begin{eqnarray}
\begin{array}{ll}
\Gamma_1=\hat{C}_L     \otimes \mathbbm{1}_L, \quad &
\Gamma_2=\hat{S}_L     \otimes \mathbbm{1}_L,       \\
\Gamma_3=\mathbbm{1}_L \otimes \hat{C}_L,           &
\Gamma_4=\mathbbm{1}_L \otimes \hat{S}_L.
\end{array}
\end{eqnarray}
From (\ref{EQ:minimal_skew_form}) we can also construct the twist-eater
configuration for the minimal symmetric twist (\ref{EQ:minimal_sym_form}) as
\begin{eqnarray}
\begin{array}{ll}
\Gamma_1=\quad\;\hat{C}_L   \otimes \mathbbm{1}_L, \quad &
\Gamma_2=\hat{S}_L\hat{C}_L \otimes \hat{C}_L,           \\
\Gamma_3=\hat{S}_L\hat{C}_L \otimes \hat{S}_L,           &
\Gamma_4=\quad\;\hat{S}_L   \otimes \mathbbm{1}_L. 
\end{array}
\end{eqnarray}
Although these forms are different only by the coordinate
transformation, they can give different results except the weak
coupling limit as seen in next section. 

Another kind of the twist we consider in this article is
\begin{eqnarray}
n_{\mu\nu}=
\left(\begin{array}{cc|cc}
 0  &  mL &  0  & 0 \\
-mL &  0  &  0  & 0 \\
\hline
 0  &  0  &  0  & mL\\
 0  &  0  & -mL & 0
\end{array}
\right),\qquad N=mL^2
\label{EQ:generic_twist}
\end{eqnarray}
with classical solution
\begin{eqnarray}
\begin{array}{ll}
\Gamma_1=\hat{C}_L     \otimes \mathbbm{1}_L \otimes \mathbbm{1}_m, \quad &
\Gamma_2=\hat{S}_L     \otimes \mathbbm{1}_L \otimes \mathbbm{1}_m,       \\
\Gamma_3=\mathbbm{1}_L \otimes \hat{C}_L     \otimes \mathbbm{1}_m,       &
\Gamma_4=\mathbbm{1}_L \otimes \hat{S}_L     \otimes \mathbbm{1}_m. 
\end{array}
\label{EQ:generic_twist-eater}
\end{eqnarray}
While we write the twist using the skew-diagonal form here, we can always
rewrite it in the symmetric form by the $SL(4, \mathbb{Z})$ transformation.
We call this twist ``generic twist'' in this article, and the minimal twists
(\ref{EQ:minimal_sym_form}) and (\ref{EQ:minimal_skew_form}) are
particular cases ($m=1$) of the generic twist. 
As is well known, the TEK model can describe the NCYM theory
\cite{Aoki:1999vr, Ambjorn:1999ts}.
Expanding the matrix model around noncommutative tori background, 
we can obtain noncommutative $U(m)$ Yang-Mills theory on fuzzy tori. 
(Note that this interpretation is possible even at finite-$N$.)
Because fuzzy torus can be used as a regularization of fuzzy
${\mathbb R}^4$, it is naively possible to give a nonperturbative
formulation of the NCYM on fuzzy ${\mathbb R}^4$ by taking a suitable
large-$N$ limit in the TEK model. (See appendix \ref{sec:NCYM} for details.)
However, we will see later it is not the case because of the
$\mathbb{Z}_N^4$ symmetry breaking. 
In the NCYM interpretation the shift and clock matrices can be regarded as
matrix realization of a fuzzy torus.
From this point of view, twist prescription (\ref{EQ:generic_twist}) provides
YM theories on $m$-coincident four-dimensional fuzzy tori.

\subsection {$\mathbb{Z}_N^4$ symmetry}\label{SEC:U(1)symmetry}

The $\mathbb{Z}_N^4$ symmetry plays a crucial role in the Eguchi-Kawai
equivalence.
Generally, the YM theory with a periodic boundary condition has
a critical size.
If we shrink the volume of the system beyond the critical size,
we encounter the center symmetry breaking, which is just the same as
the finite temperature system.
In the EK model, which is a single hyper-cubic model, the critical size 
corresponds to $\beta_c\sim 0.19$ in the lattice coupling.
In the region less than the $\beta_c$ -- the strong coupling region --
the center symmetry ${\mathbb Z}_N^4$ is maintained.
On the other hand, in the region larger than $\beta_c$
-- the weak coupling region -- the symmetry is spontaneously broken,
and then the EK equivalence does not hold.

The TEK model avoids this problem by imposing the twisted boundary
condition on the system instead of the periodic one.
In the weak coupling limit the path integral is dominated by the vacuum
configuration, which is twist-eater configurations, as we already mentioned.
These configurations are invariant under global ${\mathbb Z}_L^4$ 
transformation  
\begin{equation}
U_\mu\to e^{i\theta_\mu}U_\mu, \qquad e^{i\theta_\mu}\in {\mathbb Z}_L,
\end{equation}
which is regarded as the $U(1)^4$ symmetry in the large-$N$ limit.
As a result, $W_{TEK}({\cal C})$ is zero if $\cal C$ is an open contour
in the weak coupling limit.

A key point is that the solution for this problem is obvious only
at the classical level.
That is to say, there is no guarantee to maintain the ${\mathbb Z}_L^4$
symmetry if we take into account the quantum fluctuation.
Going away from the weak coupling limit,
the configurations fluctuate around the twist-eater.
The situation can be displayed in the eigenvalue distribution of the
link variables.
In the weak coupling limit the $N$ eigenvalues distribute regularly and
uniformly on the unit circle in the complex plane,
and then they are ${\mathbb Z}_L$ symmetric.
If we decrease $\beta$, the eigenvalues begin to fluctuate around
the location of the twist-eater.
If the fluctuation is not too large, the ${\mathbb Z}_L$ symmetric  
distribution is maintained.
However, large fluctuation can make the uniform distribution shrink to
a point, which corresponds to $U_\mu=\mathbbm{1}_N$ configuration.
In the strong coupling region the distribution is randomly uniform,
and then the symmetry is restored.

Although there is no guarantee to maintain the ${\mathbb Z}_N^4$ symmetry
in the intermediate coupling region, the 1980s numerical simulations
suggested that the symmetry was unbroken.
And this caused us to believe that the EK equivalence in the TEK model
does hold throughout the whole coupling region.

\subsection {Limiting procedure}\label{SEC:limiting_procedure}

As is well known, the scaling of the YM lattice theory behaves as
$\beta\sim\log a^{-1}$ around the weak coupling limit,
where $a$ is the lattice spacing, and which is obtained by one-loop
perturbative calculation of the renormalization group equation.
If we wish to construct the TEK model which corresponds to the YM theory
by the EK equivalence, the scaling of the TEK model should obey
that of the YM theory.
In the TEK model, the lattice size $L$ relates to $N$.
(For the twist we consider in this article, the relation is $N=mL^2$.)
Then, the YM system with fixed physical size $l=aL$ can be obtained by
the scaling
\begin{equation}
\beta\sim\log a^{-1}\sim\log N.
\label{EQ:one-loop_scaling}
\end{equation}
In order to obtain the large-$N$ limit with infinite volume,
we should increase $\beta$ slower than the scaling (\ref{EQ:one-loop_scaling}).
If it is not the case, the system shrinks to a point.

In the case of the NCYM, the scaling near the weak coupling limit
is essentially same as the YM theory, that is, $\beta\sim\log a^{-1}$.
(See appendix \ref{sec:NCYM}.)
But if we wish to make the TEK model corresponding to the NCYM,
there is a  constraint $a^2L=al=fixed$, which means that
we take a scheme in which the noncommutative parameter $\theta$ is fixed.
Then, both the continuum limit and the infinite volume limit are
simultaneously taken (double scaling limit).
Regardless of difference of the constraint, the scaling for the NCYM we should 
take is the same as that of the ordinary YM (\ref{EQ:one-loop_scaling})
by the nature of the logarithm scaling.

\section{$\mathbb{Z}_N^4$ symmetry breaking in the TEK model}
\label{SEC:breaking}

As mentioned in the previous section, the $\mathbb{Z}_N^4$ symmetry
breaking had not been observed in the older numerical simulation.
However, there are several recent reports which indicate the symmetry
breaking \cite{IO03,Teper:2006sp,Bietenholz:2006cz}.
In \cite{Teper:2006sp}, the symmetry breaking in the $D=4$ $SU(N)$ TEK model
was studied in the case of the standard twist up to $N=144=12^2$. 
The authors of \cite{Teper:2006sp} performed the Monte-Carlo
simulation starting both from a randomized configuration (``hot start'')
and from the twist-eater solution (``cold start'').
In both cases the $\mathbb{Z}_N^4$ symmetry begins to break at $N\ge 100=10^2$.
At $N=144$ the symmetry breaking and restoration patterns they observed are
\begin{eqnarray}
\begin{array}{cccccccccccl}
\mathbb{Z}_N^4 & \xrightarrow{\beta_c^H} & \mathbb{Z}_N^3 & \longrightarrow &
\mathbb{Z}_N^2 & \longrightarrow         & \mathbb{Z}_N^1 &
\multicolumn{3}{c}{\longrightarrow}      & \mathbb{Z}_N^0 \;\; &
\mbox{($N=144$, standard, hot start)},\\
\mathbb{Z}_N^4 & \longleftarrow          & \mathbb{Z}_N^3 & \longleftarrow  &
\mathbb{Z}_N^2 & \longleftarrow          & \mathbb{Z}_N^1 & \longleftarrow  &
\mathbb{Z}_N^0 & \xleftarrow{\beta_c^L}  & \mathbb{Z}_N^4 \;\; &
\mbox{($N=144$, standard, cold start)},
\end{array}
\label{EQ:breaking_pattern_sym}
\end{eqnarray}
where $\beta_c^H$ and $\beta_c^L$ are the first breaking point for the hot
start and that for cold start, respectively.
Note that although there is recovery of the symmetry for the cold start,
the symmetry remains broken for the hot start. 

In this section we show the results of the numerical simulation for
this symmetry breaking phenomena.
In order to argue about the possibility of the continuum and large-$N$
limiting procedure for this model, we mainly focus on the first breaking
point for the cold start $\beta_c^L$, which depends on $N$.\footnote{
Strictly speaking, the symmetry preserved in the weak coupling region is
not $\mathbb{Z}_N^4$ but $\mathbb{Z}_L^4$. However, $\mathbb{Z}_L^4$ is
sufficient for the Eguchi-Kawai equivalence so we do not dare to distinguish
them in this article.}

\subsection{Simulation method}

In our simulation we use the pseudo-heatbath algorithm.
The algorithm is based on \cite{Fabricius:1984wp}, and in each sweep
over-relaxation is performed five times after multiplying $SU(2)$ matrices.
The number of sweeps is $O(1000)$ for each $\beta$.
We scan the symmetry breaking on the resolution of $\Delta\beta=0.005$,
and then we always quote the value $\pm 0.0025$ as the error due to
the resolution. 
Note that the breaking points are ambiguous because the breakdown of the
$\mathbb{Z}_N^4$ symmetry is a first-order transition.
As an order parameter for detecting the $\mathbb{Z}_N^4$ breakdown,
we measure the Polyakov lines
\begin{eqnarray}
P_\mu\equiv\left\langle\left|\frac{1}{N}{\rm Tr}~U_{\mu}\right|\right\rangle.
\end{eqnarray}

\subsection{Simulation results}

\subsubsection*{Minimal symmetric twist}

\begin{figure}[t]
\begin{center}
\includegraphics[scale=0.42, viewport = 0 0 510 470, clip]
                {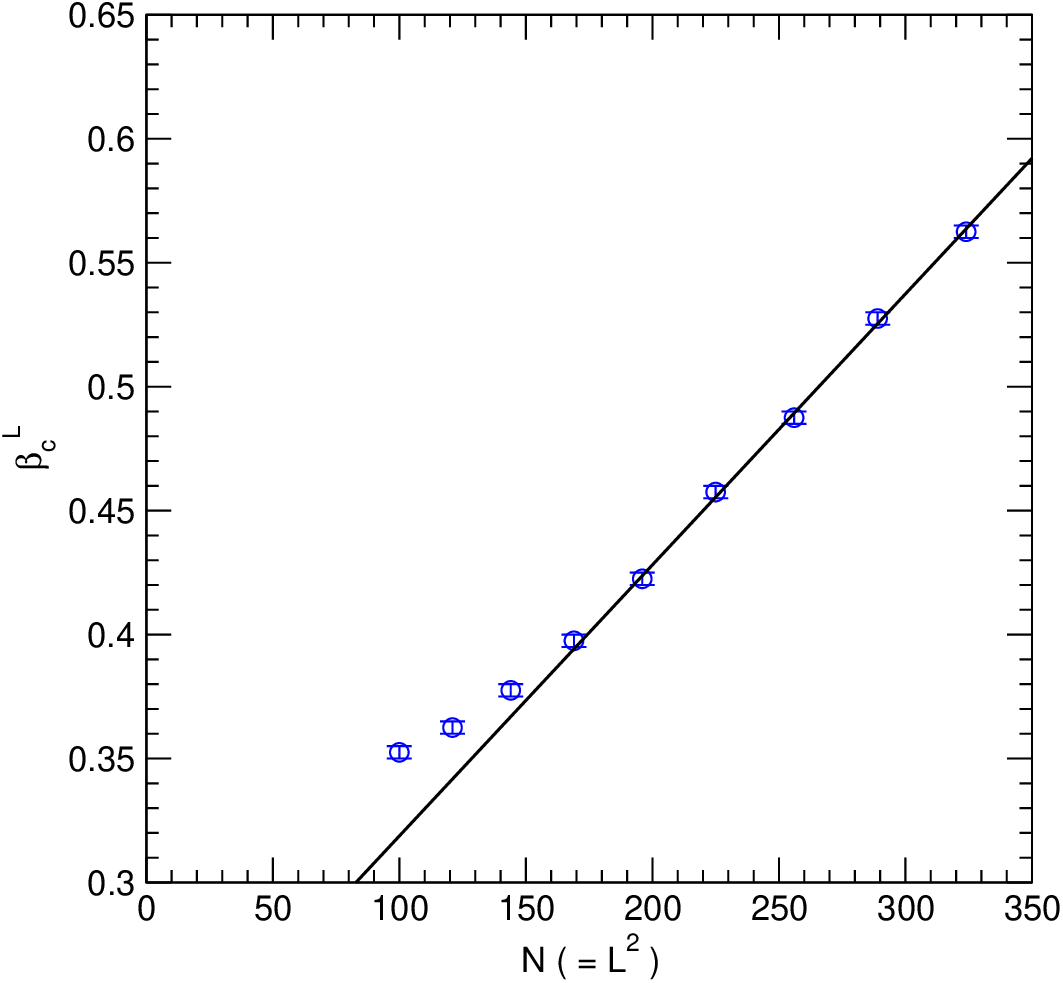}
\caption{Plot of $\beta_c^L$ versus $N$ for the minimal symmetric twist. 
Fit line is equation (\protect\ref{EQ:bc_L_minimal_sym}), which is obtained
using $N\geq169$ data.}
\label{FIG:minimal symmetric twist}
\end{center}
\end{figure}

First of all we treat the minimal symmetric twist (\ref{EQ:minimal_sym_form}).
This twist is the most standard and is also used in the paper
\cite{Teper:2006sp}. 
In our study we only investigate the first $\mathbb{Z}_N^4$ symmetry
breaking point from weak coupling limit, that is, $\beta_c^L$ for this twist.
(For more detailed information about the symmetry breaking phenomena,
see \cite{Teper:2006sp}.)
The obtained results are in table \ref{TAB:minimal symmetric twist}
and plotted in figure \ref{FIG:minimal symmetric twist}.
The symmetry breaking points and patterns
($\mathbb{Z}_N^4\xrightarrow{\beta_c^L}\mathbb{Z}_N^3$ for $N=100$;
 $\mathbb{Z}_N^4\xrightarrow{\beta_c^L}\mathbb{Z}_N^0$ for $N>100$)
are consistent with the results in \cite{Teper:2006sp} up to $N=144$.
In this work we explore the simulation for larger $N$.
From figure \ref{FIG:minimal symmetric twist} we can find clear linear
dependence of $\beta_c^L$ on $N(=L^2)$ for $N\gtrsim169$.
The fitted result in linear function using $N\geq169$ data is
\begin{equation}
\beta_c^L\sim 0.0011N+0.21.
\label{EQ:bc_L_minimal_sym}
\end{equation}
A theoretical argument for this linear behavior is discussed in
section \ref{SEC:discussions}.

\begin{table}[t]
\begin{center}
\begin{tabular}{|l|l|c||l|l|c|} \hline
$N$   & $L$  & $\beta_c^L$        & $N$   & $L$  & $\beta_c^L$        \\\hline
$100$ & $10$ & $0.3525\pm 0.0025$ & $225$ & $15$ & $0.4575\pm 0.0025$ \\\hline
$121$ & $11$ & $0.3625\pm 0.0025$ & $256$ & $16$ & $0.4875\pm 0.0025$ \\\hline
$144$ & $12$ & $0.3775\pm 0.0025$ & $289$ & $17$ & $0.5275\pm 0.0025$ \\\hline
$169$ & $13$ & $0.3975\pm 0.0025$ & $324$ & $18$ & $0.5675\pm 0.0025$ \\\hline
$196$ & $14$ & $0.4225\pm 0.0025$ & \multicolumn{3}{c|}{}             \\\hline
\end{tabular}
\end{center}
\caption{Critical lattice coupling from the weak coupling side  $\beta_c^L$
for the minimal symmetric twist.}
\label{TAB:minimal symmetric twist}
\end{table}

\subsubsection*{Minimal skew-diagonal twist}

\begin{figure}[t]
\begin{center}    
\includegraphics[scale=0.48, viewport = 0 0 650 460, clip]
                {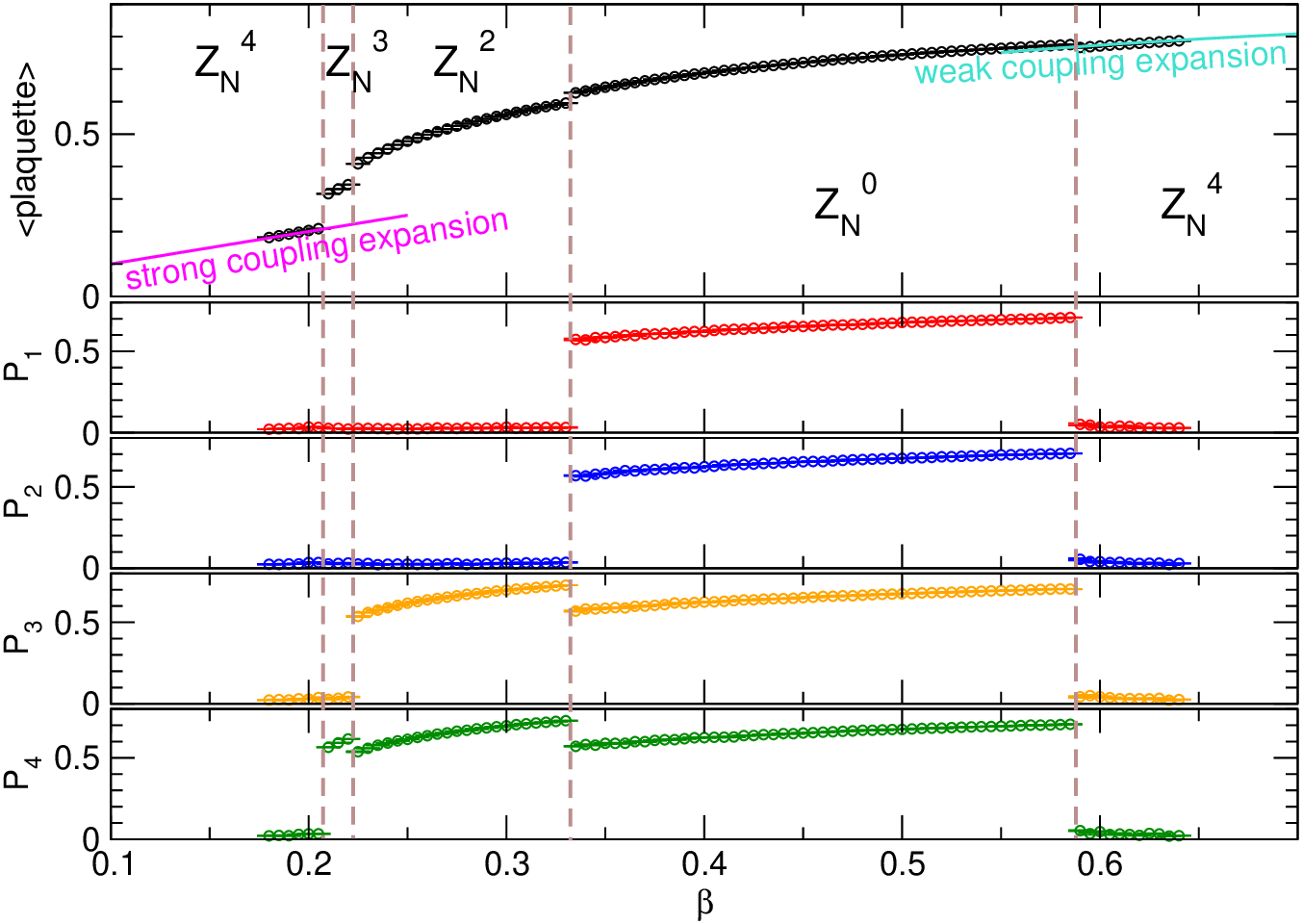}
\caption{Expectation value of the plaquette (top) and the Polyakov line
         (besides the top) versus the lattice coupling $\beta$
         for $N=100$ with the minimal skew-diagonal twist (cold start).
         As $\beta$ is decreased, the $\mathbb{Z}_N^4$ symmetry is
         broken and restored as
         $\mathbb{Z}_N^4\leftarrow\mathbb{Z}_N^3\leftarrow
          \mathbb{Z}_N^2\leftarrow\mathbb{Z}_N^0\xleftarrow{\beta_c^L}
          \mathbb{Z}_N^4$.}
\label{FIF:pattern_N100_cold-start}
\end{center}
\end{figure}

Twists can be always transformed into the skew-diagonal form by
$SL(4,\mathbb{Z})$ transformation as we mentioned in section
\ref{SEC:twist prescription}.
As it were, the minimal symmetric twist (\ref{EQ:minimal_sym_form}) is
equivalent to the minimal skew-diagonal twist (\ref{EQ:minimal_skew_form})
in the weak coupling limit.
However, both forms can represent different features by taking into account
the quantum fluctuation.
Actually, the $\mathbb{Z}_N^4$ symmetry is already broken at $N=25$.
This fact enables us to observe the $N$-dependence of the critical points  
easily.
Not only is the symmetry breaking point different from the symmetric form,
so is the breaking and restoration pattern.
Figure \ref{FIF:pattern_N100_cold-start} shows the expectation value of
the plaquette (top) and the Polyakov lines (besides the top) versus $\beta$
for the cold start at $N=100$.
For $N\geq100$ we find the $\mathbb{Z}_N^4$ symmetry breaking and
restoration pattern:
\begin{equation}
\mathbb{Z}_N^4\leftarrow\mathbb{Z}_N^3\leftarrow\mathbb{Z}_N^2\leftarrow
\mathbb{Z}_N^0\xleftarrow{\beta_c^L}\mathbb{Z}_N^4\quad
\mbox{($N=100$, minimal skew-diagonal, cold start)},
\end{equation}
which represents a difference from the symmetric form case
(\ref{EQ:breaking_pattern_sym}).
The first breaking pattern
$\mathbb{Z}_N^4\xrightarrow{\beta_c^L}\mathbb{Z}_N^0$ is,
however, the same as that in the symmetric twist.
(We note that for $N\le 81$ the first breaking pattern is 
$\mathbb{Z}_N^4\xrightarrow{\beta_c^L}\mathbb{Z}_N^2$, which resembles
the pattern $\mathbb{Z}_N^4\xrightarrow{\beta_c^L}\mathbb{Z}_N^3$ at
$N=100$ for the symmetric form \cite{Teper:2006sp}.) 

\begin{figure}[t]
\begin{center}
\begin{tabular}{lr}
\begin{minipage}{72mm}
\includegraphics[scale=0.42, viewport = 0 0 510 470, clip]
                {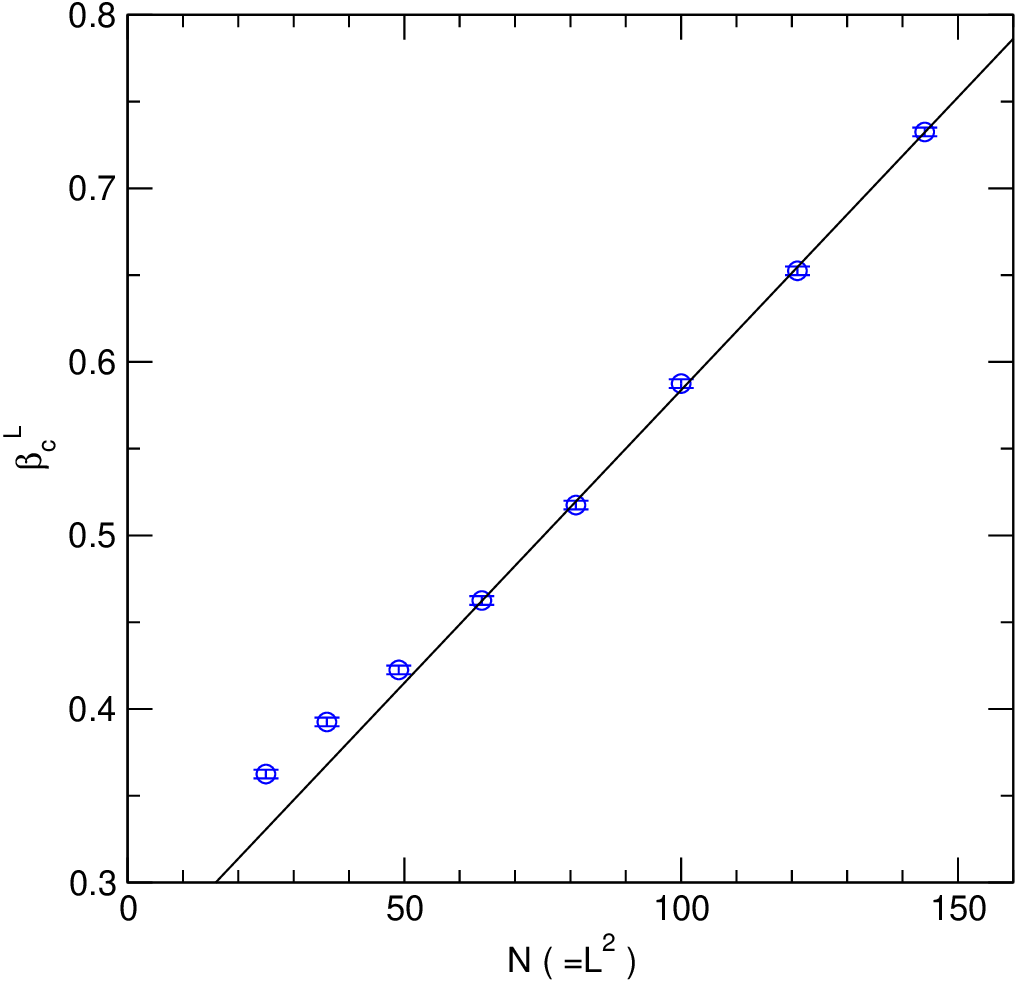}
\caption{Plot of $\beta_c^L$ versus $N$ for the minimal skew-diagonal twist.
         The fit line is equation (\protect\ref{EQ:bc_L_minimal_skew}),
         which is obtained using $N\geq64$ data.}\vspace*{13mm}
\label{FIG:m=1B_low}
\end{minipage}
&
\begin{minipage}{72mm}
\includegraphics[scale=0.42, viewport = 0 0 510 470, clip]
                {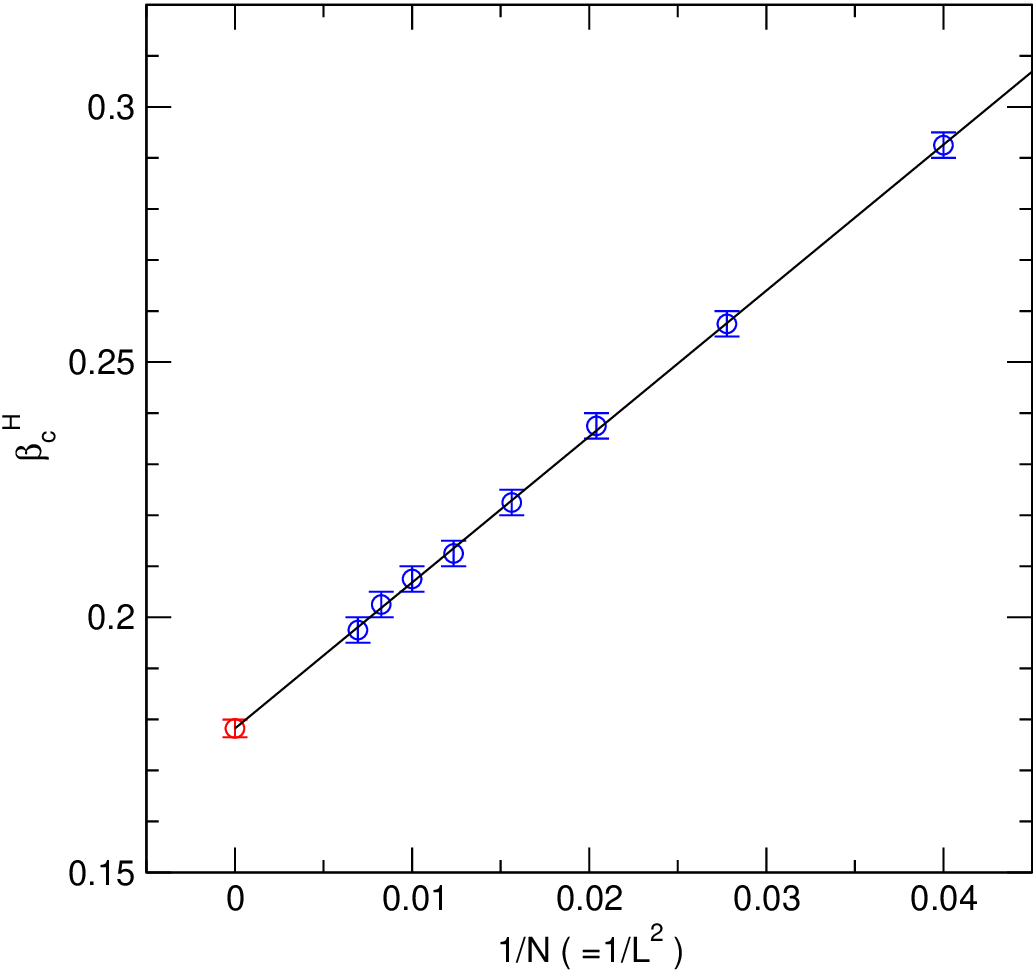}
\caption{Plot of $\beta_c^H$ versus $1/N$ for the minimal skew-diagonal twist.
         The fit line is equation (\protect\ref{EQ:bc_H_minimal_skew}),
         which is obtained using $N\geq25$ data.
         Extrapolation to $1/N=0$ gives $\beta_c^H\to 0.18$,
         which is close to the critical point in the original EK model,
         $\beta=0.19$.}
\label{FIG:m=1B_high}
\end{minipage}
\end{tabular}
\end{center}
\end{figure}

Table \ref{TAB:minimal skew-diagonal} shows the first breaking points
for the cold start $\beta_c^L$ and for the hot start $\beta_c^H$.
These data are plotted in figure \ref{FIG:m=1B_low} for $\beta_c^L$
and figure \ref{FIG:m=1B_high} for $\beta_c^H$.
Again, we find clear linear dependence on $N$ for $\beta_c^L$, as
we found for the symmetric form.
Additionally, we also find clear dependence on $1/N$ for $\beta_c^H$.
The fitted results are
\begin{eqnarray}
\beta_c^L &\sim& 0.0034N+0.25,\label{EQ:bc_L_minimal_skew}\\
\beta_c^H &\sim& \frac{2.9}{N}+0.18,\label{EQ:bc_H_minimal_skew}
\end{eqnarray}
where we used only $N\geq64$ data for $\beta_c^L$, whereas all data are
used for $\beta_c^H$.
As $N$ is increased the $\beta_c^H$ approaches a point $0.190$,
where the phase transition
$\mathbb{Z}_N^4\xrightarrow{\beta_c^H}\mathbb{Z}_N^3$ takes place
in the original EK model.
These results suggest that the quantum fluctuation is so large that
the $\mathbb{Z}_N^4$ symmetry is broken in exactly the same region as
that in the original EK model.
The lines for transitions $\beta_c^L$ and $\beta_c^H$
seem to intersect around the bulk transition point $\beta_c^B\sim0.35$,
which corresponds to $N\sim 20$ for the twist considered here.
For smaller values than $N\sim 20$, we did not
observe a signal of breakdown of the $\mathbb{Z}_N^4$ symmetry.

\begin{table}[t]
\begin{center}
\begin{tabular}{|l|l|c|c||l|l|c|c|} \hline
$N$   & $L$  & $\beta_c^H$        & $\beta_c^L$        &
$N$   & $L$  & $\beta_c^H$        & $\beta_c^L$        \\\hline
$9$   & $3$  & -                  &-                   &
$64$  & $8$  & $0.2225\pm 0.0025$ & $0.4625\pm 0.0025$ \\\hline
$16$  & $4$  & -                  &-                   &
$81$  & $9$  & $0.2125\pm 0.0025$ & $0.5175\pm 0.0025$ \\\hline
$25$  & $5$  & $0.2925\pm 0.0025$ & $0.3625\pm 0.0025$ &
$100$ & $10$ & $0.2075\pm 0.0025$ & $0.5875\pm 0.0025$ \\\hline
$36$  & $6$  & $0.2575\pm 0.0025$ & $0.3925\pm 0.0025$ &
$121$ & $11$ & $0.2025\pm 0.0025$ & $0.6525\pm 0.0025$ \\\hline
$49$  & $7$  & $0.2375\pm 0.0025$ & $0.4225\pm 0.0025$ &
$144$ & $12$ & $0.1975\pm 0.0025$ & $0.7325\pm 0.0025$ \\\hline
\end{tabular}
\caption{Critical lattice coupling from the weak coupling side
         $\beta_c^L$ and from strong coupling side $\beta_c^H$
         for the minimal skew-diagonal twist $(m=1)$.}
\label{TAB:minimal skew-diagonal}
\end{center}
\end{table}

\subsubsection*{Generic skew-diagonal twist}

Here, we show the numerical result of the generic twist
(\ref{EQ:generic_twist}).
For this twist we use the skew-diagonal form because the
$\mathbb{Z}_N^4$ symmetry breaking occurs at smaller $N$ than
that in the symmetric form, which makes our investigation much easier.
 
\begin{figure}[t]
\begin{center}
\begin{tabular}{lr}
\hspace*{-3mm}
\begin{minipage}{72mm}
\includegraphics[scale=0.42, viewport = 0 0 510 470, clip]
                {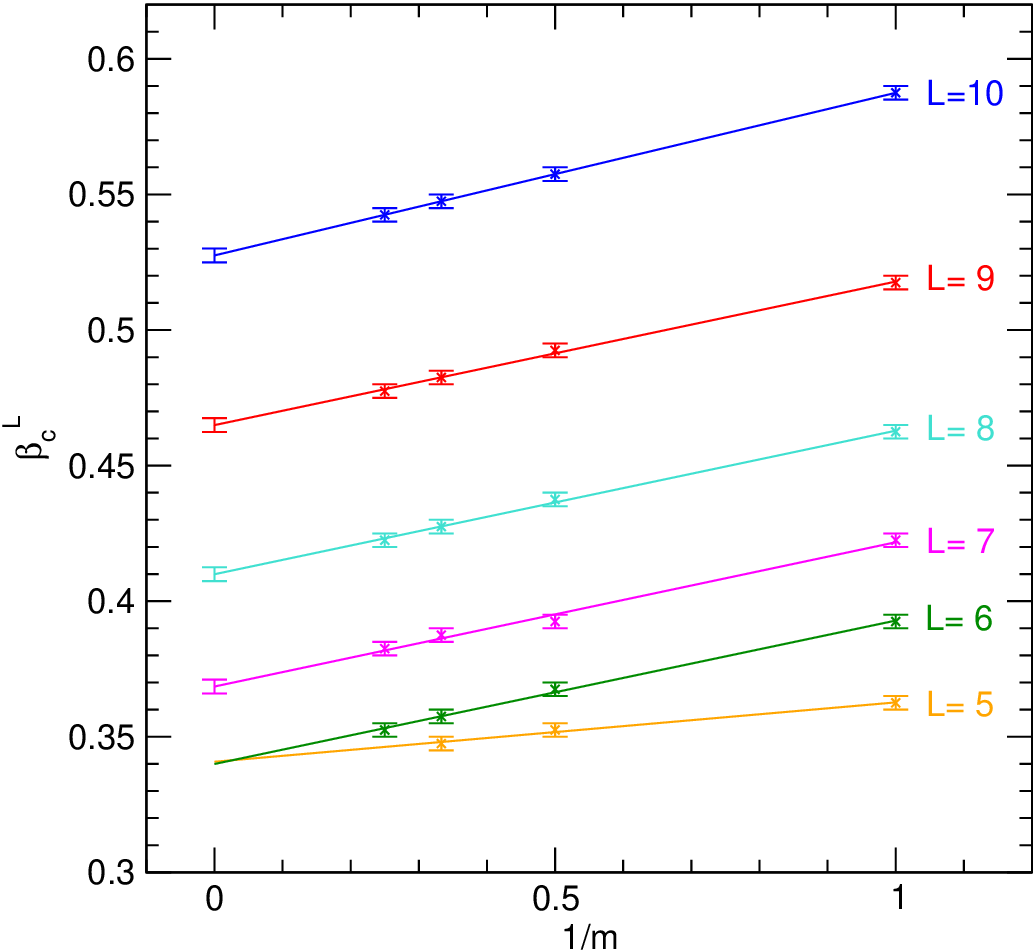}
\caption{$\beta_c^L$ versus $1/m$ for $L=5,\cdots,10$.
         $\beta_c^L$ for $m=\infty$ is evaluated by extrapolating
         these data with straight line.}\vspace*{+2mm}
\label{FIG:bc_L_mall_skew1}
\end{minipage}
&
\begin{minipage}{72mm}
\includegraphics[scale=0.42, viewport = 0 0 510 470, clip]
                {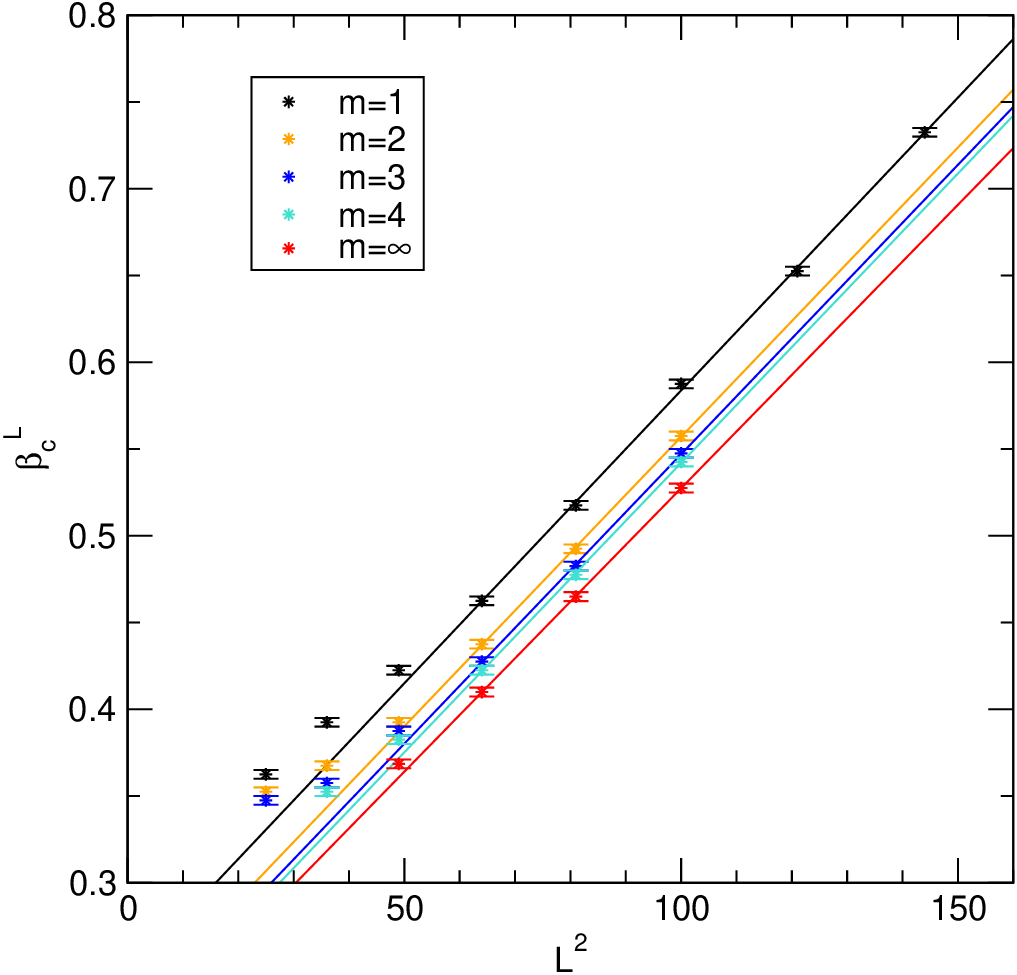}
\caption{Scaling of $\beta_c^L$ for $m=1,2,3$ and $4$.
         We also include $\beta_c^L$ for $m=\infty$, which is obtained
         by an extrapolation shown in figure
         \protect\ref{FIG:bc_L_mall_skew1}.}
\label{FIG:bc_L_mall_skew2}
\end{minipage}
\end{tabular}
\end{center}
\end{figure}

We measure $\beta_c^L$ for this twist up to $m=4$.
Table \ref{TAB:m=234} shows the $\beta_c^L$ for $m=2, 3, 4$ and
that for $m=1$ is presented in table \ref{TAB:minimal skew-diagonal}.
These data are plotted in figure \ref{FIG:bc_L_mall_skew1}.
From this figure we can find that the $\beta_c^L$ for each $L$ are reduced  
as we increase $m$, and the dependence is linear in $1/m$.
The data at $1/m=0$ in this plot are linearly extrapolated values.
An interesting point is the behavior for the case $L=5$.
While the $\mathbb{Z}_N^4$ symmetry breaking is observed for $m=1, 2$ and $3$,
it is not seen for $m=4$ because the $\beta_c^L$ reaches the bulk transition
point $\beta_c^B\sim0.35$ by increasing $m$.
Figure \ref{FIG:bc_L_mall_skew2} represents the same data in
figure \ref{FIG:bc_L_mall_skew1}, but the horizon axis is $L^2$.
As we have seen in the $m=1$ case, the data for $L\geq8$ are well fitted
by the linear function of $L^2$ for each $m$.
From these figures, we find that the data for $L\geq8$ are well
fitted globally by a function:
\begin{equation}
\beta_c^L\sim0.0034L^2+\frac{0.060}{m}+0.19.
\label{EQ:bc_L_generic_skew}
\end{equation}

\begin{table}[t]
\begin{center}
\begin{tabular}{|l||l|c||l|c||l|c|} \hline
     & \multicolumn{2}{c||}{$m=2$} & \multicolumn{2}{c||}{$m=3$} &
       \multicolumn{2}{c| }{$m=4$}   \\\hline
$L$  & $N$   & $\beta_c^L$         & $N$   & $\beta_c^L$         &
       $N$   & $\beta_c^L$           \\\hline
$5$  & $50$  & $0.3525\pm0.0025$   & $75$  & $0.3475\pm0.0025$   &
       -     & -                     \\\hline
$6$  & $72$  & $0.3675\pm0.0025$   & $108$ & $0.3575\pm0.0025$   &
       $144$ & $0.3525\pm0.0025$     \\\hline
$7$  & $98$  & $0.3925\pm0.0025$   & $147$ & $0.3875\pm0.0025$   &
       $196$ & $0.3825\pm0.0025$     \\\hline
$8$  & $128$ & $0.4375\pm0.0025$   & $192$ & $0.4275\pm0.0025$   &
       $256$ & $0.4225\pm0.0025$     \\\hline
$9$  & $162$ & $0.4925\pm0.0025$   & $243$ & $0.4825\pm0.0025$   &
       $324$ & $0.4775\pm0.0025$     \\\hline
$10$ & $200$ & $0.5575\pm0.0025$   & $300$ & $0.5475\pm0.0025$   &
       $400$ & $0.5425\pm0.0025$     \\\hline
\end{tabular}
\caption{$\beta_c^L$ for the generic skew-diagonal twist ($m=2, 3, 4$).
         See also Tab. \protect\ref{TAB:minimal skew-diagonal} for $m=1$.}
\label{TAB:m=234}
\end{center}
\end{table}
 
\section{Discussions}\label{SEC:discussions}

In this section we discuss the numerical results obtained
in the previous section and the validity of taking the large-$N$ and
continuum limit for this model.

\subsection{Theoretical estimation of the $\mathbb{Z}_N^4$
            symmetry breaking point}
\label{SEC:theoretical_estimation}            

In the previous section we showed our numerical results. In particular,
we elaborately investigated $\beta_c^L$, the first $\mathbb{Z}_N^4$
breaking point from the cold start.
From our investigation, we found the clear linear behavior like
(\ref{EQ:bc_L_minimal_sym}), (\ref{EQ:bc_L_minimal_skew}) and
(\ref{EQ:bc_L_generic_skew}).
These behaviors can be obtained through the following consideration.

\subsubsection*{Energy difference between twist-eater $\Gamma_{\mu}$ and
                identity $\mathbbm{1}_N$ configurations}

We simply assume that the $\mathbb{Z}_N^4$ breaking is a transition
from twist-eater phase $U_{\mu}=\Gamma_{\mu}$ to identity configuration
phase $U_{\mu}=\mathbbm{1}_N$.
For plainness, we consider
$\mathbb{Z}_N^4\xrightarrow{\beta_c^L}\mathbb{Z}_N^0$ type breaking here.
Of course we can treat
$\mathbb{Z}_N^4\xrightarrow{\beta_c^L}\mathbb{Z}_N^3\xrightarrow{\beta_c^L}
 \mathbb{Z}_N^2\xrightarrow{\beta_c^L}\mathbb{Z}_N^1\xrightarrow{\beta_c^L}
 \mathbb{Z}_N^0$ (cascade) type breaking at a $\beta_c^L$,
but the obtained behavior is not different from the former type.
Firstly, we focus on the classical energy difference between these
configurations. 
The energy 
 can be easily calculated from the action
(\ref{EQ:TEK_action}) as
\begin{eqnarray}
\Delta S
&=&S_{TEK}[U_{\mu}=\mathbbm{1}_N]-S_{TEK}[U_{\mu}=\Gamma_{\mu}]\nonumber\\
&=&\beta N^2\sum_{\mu\neq\nu}
   \left\{ 1-\cos\left(\frac{2\pi n_{\mu\nu}}{N}\right)\right\}
   \simeq 2\pi^2\beta\sum_{\mu\neq\nu}n_{\mu\nu}^2. 
\label{EQ:potential_difference}
\end{eqnarray}
For the generic twist, it becomes
\begin{equation}
\Delta S=
\begin{cases}
24\pi^2\beta m^2L^2\qquad & \mbox{(symmetric form)},\\
8\pi^2\beta m^2L^2\qquad  & \mbox{(skew-diagonal form)}.
\label{EQ:potential_difference_generic}
\end{cases}
\end{equation}
Note that the symmetric form is roughly three times more stable than
the skew-diagonal form if both twists have equal quantum fluctuations.
This is the reason that the $\mathbb{Z}_N^4$ symmetry breaking for
the skew-diagonal form can occur at quite smaller $N$ than
that for the symmetric form, as is observed in our simulation. 

\subsubsection*{Quantum fluctuations and symmetry breaking}
\label{SSEC:fluctuation}

Going away from the weak coupling limit, the system has quantum fluctuations.
Here we naively expect that the $\mathbb{Z}_N^4$ symmetry is broken
if the fluctuation around twist-eater configuration exceeds
the energy difference $\Delta S$.
Because the system describes $O(N^2)$ interacting gluons,
it is natural to assume that their quantum fluctuations provide
$O(N^2)$ value to the effective action.
For the generic twist the quantum fluctuation is $O(m^2L^4)$.
Combined with the fact (\ref{EQ:potential_difference_generic}),
we can estimate the critical point $\beta_c^L$ as
\begin{equation}
\beta_c^L\sim L^2,\label{EQ:linear_behavior}
\end{equation}
which is consistent with the numerical results (\ref{EQ:bc_L_minimal_sym}), 
(\ref{EQ:bc_L_minimal_skew}) and (\ref{EQ:bc_L_generic_skew}).
In addition we can explain the difference of the coefficient of $L^2$
in (\ref{EQ:bc_L_minimal_sym}), (\ref{EQ:bc_L_minimal_skew}) and
(\ref{EQ:bc_L_generic_skew}) between the symmetric and the skew-diagonal form,
which is roughly three times different, by the factor in
(\ref{EQ:potential_difference_generic}).

Although the above crude estimation reproduces the linear $L^2$ behavior
of $\beta_c^L$, we cannot explain the dependence on $m$.
To catch the behavior completely, we need to make the discussion
more sophisticated.
However, we do not pursue this issue here because the $m$ dependence
can be negligible at the larger $N$.

This argument can be applied for other twist prescriptions like
taking the twist phase as $\exp(i\pi(L+1)/L)$,
which is usually used for describing noncommutative spaces.
(See appendix \ref{sec:NCYM}.)

\subsection{Continuum and large-$N$ limit}

We have shown that the linear $L^2$ dependence of the critical point
$\beta_c^L$ could be explained by the theoretical discussion in this section.
While our simulation is restricted in the small $N$ region,
we confirm that the behavior must continue to $N=\infty$ by combining with
the discussion.
Then the EK equivalence is valid only in the region $\beta >\beta_c^L\sim N$
even in the weak coupling limit and the large-$N$ limit.
As we mentioned in the section \ref{SEC:limiting_procedure},
both the ordinary YM with fixed physical volume and the NCYM theory with
fixed noncommutative parameter have essentially logarithm scaling
(\ref{EQ:one-loop_scaling}) near the weak coupling limit.
Then, because $\beta_c^L$ grows faster than the logarithm,
the EK equivalence does not hold in the continuum limit.

\section{Conclusions}

In order to study the nonperturbative nature of the large-$N$ gauge theory
by lattice simulations, the large-$N$ reduction is very useful
property for saving the computational effort.
In this paper, we studied the phase structure of the TEK model,
which has been a major way to realize the large-$N$ reduction.
Contrary to the naive hope in old days, at least in ordinary twist
prescriptions as investigated in this paper,
the $\mathbb{Z}_N^4$ symmetry is broken even
in the weak coupling region and hence a continuum limit as the planar
gauge theory cannot be described by the TEK model.
For the NCYM, the situation is the same.
We can also consider a lot of variation for the twist prescription
and the combination of reduced and non-reduced dimension.
For example, in \cite{Guralnik:2002ru,Bietenholz:2006cz}, 
four-dimensional model with two commutative and two noncommutative 
directions was studied using two-dimensional lattice action. 
However, the $\mathbb{Z}_N^4$ symmetry is broken also in this model,
and hence we cannot take a naive continuum limit.


Another way for the reduction is the QEK model, in which the eigenvalues
of the link variables are quenched.
The QEK model might have no problem in principle,
but its computational cost is larger than that of the TEK model.
Although the TEK and QEK model are reduced models to a single hyper-cube, 
recent studies deviate from them.
The contemporary method might be the partial reduction \cite{Narayanan:2003fc}.
This work showed that the large-$N$ reduction is valid above some
critical physical size $l_c$.
This means that for a lattice size $L$ the reduction holds below some
lattice coupling $\beta(L)$.
In order to take continuum limits we should avoid the bulk transition
point $\beta_c^B$, causing the condition $\beta_c^B<\beta(L)$ to be necessary.
That is, there is a lower limitation for the lattice size $L_c$ for the
 continuum reduction.
In addition, the twist prescription is also applicable to the partial
reduction \cite{GonzalezArroyo:2005dz}.
Due to the twisted boundary condition, the lower limitation $L_c$ can
be reduced.
Therefore, combination of the twist prescription and the partial
reduction would be quite efficient in the current situation.

Note also that NCYM on fuzzy ${\mathbb R}^4$ could be realized
by using TEK with quotient conditions \cite{Ambjorn:1999ts} which
give a periodic condition to eigenvalues
and hence quantum fluctuation is suppressed.
Further study in this direction would be important.
\acknowledgments

The numerical computations in this work were in part carried out at
the Yukawa Institute Computer Facility. 
The authors would like to thank Sinya Aoki, Hikaru Kawai, Jun Nishimura,
Masanori Okawa, Yoshiaki Susaki, Hiroshi Suzuki and Shinichiro Yamato
for stimulating discussions and comments.
M.~H. was supported by Special Postdoctoral Researchers Program at RIKEN.
T.~H. would like to thank the Japan Society for the Promotion of Science
for financial support. 

\appendix

\section{Double scaling limit as the noncommutative Yang-Mills theory}
\label{sec:NCYM}

The TEK model can be used to formulate gauge theories
on noncommutative spaces nonperturbatively
\cite{Aoki:1999vr,Ambjorn:1999ts,Connes:1997cr}. 
In this appendix, we give a review for the construction of the NCYM
from the TEK model \cite{Aoki:1999vr}, a discussion for the scaling
and some supplemental comments for our analysis.

By taking $U_\mu=e^{iaA_\mu}$, where $a$ corresponds to the lattice
spacing, and expanding the action of the TEK model (\ref{EQ:TEK_action}),
we have its continuum version
\begin{equation}
S_{TEK, continuum}=-\frac{1}{4g^2}\sum_{\mu\neq\nu}
         {\rm Tr}\left([A_\mu,A_\nu]-i\theta_{\mu\nu}\right)^2
\label{EQ:continuum_TEK}
\end{equation}
up to higher order terms in $a$, where 
\begin{equation}
\theta_{\mu\nu}=\frac{2\pi n_{\mu\nu}}{Na^2},\quad
\frac{1}{2g^2}=a^4 \beta N.   
\end{equation}
Then, by expanding the action around a classical solution of
(\ref{EQ:continuum_TEK})
\begin{equation}
A_\mu^{(0)}=\hat{p}_\mu, \qquad
[\hat{p}_\mu,\hat{p}_\nu]=i\theta_{\mu\nu},   
\end{equation}
we obtain the $U(1)$ NCYM on fuzzy ${\mathbb R}^4$ as follows. 
Let us define the ``noncommutative coordinate''
$\hat{x}^\mu=\left(\theta^{-1}\right)^{\mu\nu}\hat{p}_\nu$.
Then we have 
\begin{equation}
[\hat{x}^\mu,\hat{x}^\nu]=-i(\theta^{-1})^{\mu\nu}. 
\end{equation}
This commutation relation is the same as that of coordinate on fuzzy
${\mathbb R}^4$ with noncommutativity parameter $\theta$, and hence
functions of $\hat{x}$ can be mapped to functions on fuzzy ${\mathbb R}^4$.
More precisely, we have the following mapping rule:  
\begin{eqnarray}
\begin{array}{ccc}
f(\hat{x})=\sum_k\tilde{f}(k)e^{ik\hat{x}} & \leftrightarrow &
f(x)=\sum_k\tilde{f}(k)e^{ikx},\\
f(\hat{x})g(\hat{x})                       & \leftrightarrow &
f(x)\star g(x),\\
i[\hat{p}_{\mu},\ \cdot\ ]                 & \leftrightarrow &
\partial_{\mu},\\
{\rm Tr}                                   & \leftrightarrow &
\frac{\sqrt{\det\theta}}{4\pi^2}\int d^4x,  
\end{array}
\end{eqnarray}
where $\star$ represents the noncommutative star product, 
\begin{equation}
f(x)\star g(x)=
f(x)\exp\left(-\frac{i}{2}\overset{\leftarrow}{\partial}_{\mu}
(\theta^{-1})^{\mu\nu}\overset{\rightarrow}{\partial}_{\nu}\right) g(x),
\end{equation}
and we obtain $U(1)$ NCYM action
\begin{equation}
S_{U(1)NC}=-\frac{1}{4g_{NC}^2}\int d^4x ~F_{\mu\nu}\star F_{\mu\nu}
\end{equation}
with coupling constant 
\begin{equation}
g_{NC}^2=4\pi^2g^2/\sqrt{\det\theta}.
\end{equation}
In the same way, by expanding the action (\ref{EQ:continuum_TEK}) around
$A_\mu^{(0)}=\hat{p}_\mu\otimes\mathbbm{1}_m$, $U(m)$ NCYM can be obtained. 
From (\ref{EQ:generic_twist-eater}), it is apparent that
the generic twist gives the $U(m)$ NCYM. 
Intuitively, the vacuum configuration (\ref{EQ:generic_twist-eater}) 
describes $m$-coincident fuzzy tori and fuzzy ${\mathbb R}^4$ is 
realized as a tangent space. 

In order to keep the noncommutative scale $\theta$ finite, we should take the
double scaling limit with
\begin{eqnarray}
a^{-1}\sim\sqrt{L}\sim N^{1/4}. 
\end{eqnarray}
One-loop beta function for $U(m)$ NCYM is given by
\cite{Minwalla:1999px}\footnote{
Renormalizability of the NCYM is a delicate problem.
For example, see \cite{Bichl:2001cq}, in which the renormalizability is
discussed by a perturbation expansion.}
\begin{eqnarray}
\beta_{\rm 1-loop}(g_{NC})=
-\frac{2}{(4\pi)^2}\frac{11}{3}mg_{NC}^3+O(g_{NC}^5). 
\end{eqnarray}
Therefore, the 't~Hooft coupling $\beta$ scales as
\begin{eqnarray}
\beta \sim \frac{1}{g_{NC}^2} \sim \log N. 
\end{eqnarray}
Then, the scaling we should take for the NCYM is just the same
as that for the ordinary YM, and $\mathbb{Z}_N^D$ symmetry is broken
in the scaling limit.
Therefore, fuzzy torus crunches to a point
and hence the fuzzy ${\mathbb R}^4$ cannot be realized\footnote{
Although the fuzzy ${\mathbb R}^4$ cannot be realized,
another kind of noncommutative space with finite physical volume
may exist. In the case of a four-dimensional model
with two commutative and two noncommutative directions,
such a limit was found numerically \cite{Bietenholz:2006cz}.
}.

Of course, we can also use other twist prescriptions.
In order to make the periodicity of the discretized fuzzy torus 
correct, we usually take the twist as $\exp(i\pi(L+1)/L)$
\cite{Ambjorn:1999ts}.
Regardless of the difference of the twist,
the conclusion might not be altered.
Here we repeat the discussion in section
\ref{SEC:theoretical_estimation}.
In this case, the $\mathbb{Z}_N^4$ is likely to break down to
${\mathbb Z}_2^4$.
The difference between potentials in twist-eater and
${\mathbb Z}_2^4$-preserving configurations is
\begin{equation}
\Delta S\sim\beta N^2\left\{
1-\cos\left(\frac{\pi}{L}\right)\right\}\sim\beta m^2 L^2,
\end{equation}
which is the same order as (\ref{EQ:potential_difference}).
Then the behavior of the critical point $\beta_c^L$
(\ref{EQ:linear_behavior}) is not changed.


\end{document}